\documentclass[fleqn,usenatbib]{mnras}

\usepackage{newtxtext,newtxmath}

\usepackage[T1]{fontenc}
\usepackage{ae,aecompl}

\usepackage{graphicx}	
\usepackage{amsmath}	
\usepackage{amssymb}	
\usepackage{multirow}   
\usepackage{siunitx}    

\hypersetup{draft}

\title[Spiral structure in EC21178--5417]{Analysis of spiral structure in the accretion disc of the novalike Cataclysmic Variable EC21178--5417}

\author[R. Ruiz-Carmona et al.]{
R. Ruiz-Carmona,$^{1}$\thanks{E-mail: r.ruizcarmona@astro.ru.nl}
Z. N. Khangale,$^{2,3}$
P. A. Woudt,$^{2}$
P. J. Groot$^{1,2,3,4}$
\\
$^{1}$Department of Astrophysics/IMAPP, Radboud University, P.O. Box 9010,
6500 GL Nijmegen, The Netherlands\\
$^{2}$ Department of Astronomy, University of Cape Town, Private Bag X3,
Rondebosch, 7701, South Africa \\
$^{3}$ South African Astronomical Observatory, P.O. Box 9, Observatory
7935, Cape Town, South Africa \\
$^{4}$ The Inter-University Institute for Data Intensive Astronomy,
University of Cape Town, Private Bag X3, Rondebosch, 7701, South Africa \\
}

\date{Accepted 2019 October 4. Received 2019 September 27; in original form 2019 March 7.}

\pubyear{2019}

\begin{document}
\label{firstpage}
\pagerange{\pageref{firstpage}--\pageref{lastpage}}
\maketitle

\begin{abstract}
We present an extensive Doppler tomography study of the eclipsing novalike EC21178--5417, which exhibits the classic accretion disc signature in the form of double-peak emission lines in its spectrum. Doppler tomograms confirm the presence of a strong, two-armed spiral pattern visible in the majority of the spectral lines studied. This makes EC21178--5417 one of the very few novalikes that show spiral structure in their discs. We also report night-to-night changes in the position and relative strength of the spiral arms, revealing fluctuations on the conditions in the accretion disc.
\end{abstract}




\section{Introduction}

Cataclysmic variables (CVs) are stellar binary systems that contain a white dwarf and a low-mass, main-sequence star that is transferring mass via Roche-lobe overflow \citep{1995CAS....28.....W}. In systems where the white dwarf primary is non- or weakly-magnetic the gas from the secondary star forms an accretion disc around the primary. These discs are susceptible to magneto-rotational and thermal instabilities \citep{1974PASJ...26..429O}, due to coupling of sufficiently ionised gas with a (disc) magnetic field in a differentially rotating gas flow \citep{1991ApJ...376..214B}. Dwarf novae are a subtype of CVs, characterised by secularly low mass-transfer rates ($\dot{M}\leq10^{-8} M_\odot\, \rm{yr}^{-1}$), in which disc instabilities induce cycling between a cold, quiescent state and a hot, outbursting state. In the novalike subclass, the mass transfer rate from the secondary lies above the dwarf nova regime, and due to the subsequent higher disc temperatures (e.g. \citealt{1993Natur.362..518R} on UX\,UMa, and \citealt{2004A&A...417..283G} on RW\,Tri), the accretion discs are stabilized against the dwarf nova instability mechanism. A detailed study of the structure of novalike and dwarf nova discs will therefore contribute to our understanding of the physics of accretion disc in various mass-transfer rates regimes. 

Since accretion discs in CVs cannot be resolved using current techniques, many observational studies rely on indirect imaging techniques, such as Doppler tomography \citep{1988MNRAS.235..269M}. Doppler tomography uses the information in the phase-resolved profiles of emission lines to produce a two-dimensional image of the accretion disc in velocity space, virtually resolving the disc on microarcsecond scales. Doppler tomograms of most dwarf novae reveal a distinctive ring-like accretion disc structure, and the majority show evidence of emission from the secondary star and from the region where the gas stream hits the accretion disc, i.e. the bright spot (see \citealt{2001LNP...573...B}, \citealt{2005Ap&SS.296..403M}, \citealt{2012MmSAI..83..570E} and \citealt{2016ASSL..439..195M} for comprehensive and pedagogical reviews). For a number of dwarf novae, especially during an outburst, Doppler tomograms expose strong asymmetries in their discs, e.g. IP\,Peg \citep{1997MNRAS.290L..28S}, EX\,Dra \citep{2000A&A...356L..33J}, U\,Gem \citep{2001ApJ...551L..89G}, WZ\,Sge \citep{2002PASJ...54L...7B}. These asymmetries are interpreted as spiral density waves; tidal perturbations induced by the secondary star that propagate as waves, and can steepen into shocks (\citealt{1986MNRAS.219...75S}, \citealt{1987A&A...184..173S}). Spiral shocks are important to understand the physics of accretion discs, as they can enhance the effective viscosity and drive angular momentum transport \citep{1986MNRAS.219...75S} at rates sensitive to the disc temperature \citep{1987A&A...184..173S}. Spiral density waves appear more openly wound when discs are hot \citep{1999MNRAS.307...99S}, which favours detectability during the outburst state.

There are critical views on the interpretation of asymmetric emission regions in tomograms as spiral density waves motivated by a number of issues such as the uneven relative flux of the spiral arms observed in tomograms (e.g. \citealt{1999MNRAS.306..348H}, \citealt{2001ApJ...551L..89G}), or the unrealistic high temperatures necessary for numerical simulations to produce openly wound, detectable spiral density waves \citep{1998MNRAS.295L..11G}. Alternatives explanations for the emission observed in tomograms, such as converging, tidally distorted gas orbits \citep{2001AcA....51..295S} or tidally thickened, irradiated sectors of the disc \citep{2002MNRAS.330..937O}, have been put forward.

Despite the fact that novalikes are often regarded as dwarf novae that are always in outburst state \citep{2010ApJ...719.1932N}, their spectroscopic characteristics have been much more challenging to understand (\citealt{1987ApJS...65..451H}, \citealt{1994ApJS...93..519K}). Many novalike systems show emission line profiles that are not following the classic double-peaked profile of an accretion disc, although eclipse mapping studies do show the presence of a disc (e.g. \citealt{1995ApJ...448..395B}). Doppler maps of the majority of novalikes do not display signs of an accretion disc, but rather emission appears concentrated towards the centre of the tomogram or slightly shifted towards the lower left quadrant \citep{2001LNP...573...B}. This dichotomy in appearance strongly hints at non-Keplerian flows dominating the gas in the line emitting region. 

Spiral structures in novalikes are not immediately expected, V347\,Pup is an exception among novalike systems. V347\,Pup is an eclipsing, 13th magnitude novalike \citep{1990ApJ...355..617B}, that showed spiral structure only in a combined Balmer tomogram \citep{1998MNRAS.299..545S}. The presence and strength of this spiral pattern appears to be time-dependent, as \citet{2001LNP...573...45S} and later \citet{2005MNRAS.357..881T} reported that the spiral structure was not visible in Balmer lines a few years later, but were still distinguishable in some He \textsc{i} lines. Spiral structure has been also discovered in the novalike V3885 Sgr \citep{2005MNRAS.363..285H}, most convincingly in H$\beta$ and H$\gamma$, and UX\,UMa \citep{2011MNRAS.410..963N}, visible in H$\alpha$ only. 

EC21178--5417, hereafter EC2117--54, is an eclipsing, 13.7th magnitude novalike discovered in the Edinburgh-Cape survey of blue objects \citep{1997MNRAS.287..848S}. \citet{2003MNRAS.344.1193W} determined the orbital period and found the system to be prone to dwarf-nova oscillations and quasi-periodic oscillations. \citet{Khangale:2013uq} presented a detailed study of phase-resolved spectroscopic data on EC2117--54, and computed low resolution Doppler tomograms of H$\alpha$, H$\beta$ and He\,\textsc{ii}\,4686 where clear asymmetries are visible in the accretion disc.

In this paper, we present a tomographic study of a series of Balmer, He\,\textsc{i} and He\,\textsc{ii} spectral lines of EC2117--54 that allows us to confirm the presence, and study the evolution of the spiral structure in many lines over a week's worth of data. We describe our observations in Section~\ref{sec:obs} and we describe our analysis in Section~\ref{sec:analysis}. We present our tomograms and characterise the spiral structure in Section~\ref{sec:results}. We discuss and summarise our findings in Section~\ref{sec:discussion} and Section~\ref{sec:conclusions}.


\section{Observations}
\label{sec:obs} 

We observed the system EC2117--54 ($\alpha = 21^h 21^m 26.63^s,\; \delta=-54^{\circ} 04' 34.7''$) on the week of October 5--11, 2016. We obtained time-resolved spectroscopic data on six nights with the spectrograph SpUpNIC \citep{2016SPIE.9908E..27C}, mounted at the Cassegrain focus of the 1.9-meter telescope at the Sutherland station of the South African Astronomical Observatory (SAAO) in South Africa. 

A summary of the acquired spectroscopic data is provided in Table~\ref{tab:data}. All spectra were corrected for bias level and flat fielding, and optimally extracted in the usual fashion with {\sc iraf} and additionally written code in {\sc p}y{\sc raf}. Lamp arcs were taken every two science exposures, and used to calibrate all spectra in wavelength. The spectrophotometric standard star LTT 7379 (\citealt{1992PASP..104..533H}, \citealt{1994PASP..106..566H}) was also observed, and used to flux calibrate all spectra. 

Additionally, in order to constrain the time of eclipse, we observed EC2117--54 on July 19, 2016 with the SAAO 1.9-m telescope equipped with the Sutherland High-speed Optical Camera (SHOC, \citealt{2011epsc.conf.1173G,2013PASP..125..976C}), which utilizes an Andor iXon888 CCD cameras with 1024 $\times$ 1024 pixels. The SHOC camera was used in frame-transfer mode with a clear filter, and an exposure time of 1 s. The resulting data cube was reduced using the SHOC-pipeline described in \citet{2013PASP..125..976C}.

 \begin{table*}
 	\centering
    \renewcommand*{\arraystretch}{1.3}
 	\caption{Summary of spectroscopic observations of EC2117--54. The observations on 2016-10-06 do not span a sufficient portion of the orbital period to compute Doppler tomograms.}
 	\label{tab:data}
 	\begin{tabular}{lcc c c cc ccc} 
 		\hline
        \multirow{4}{1.3cm}{\centering{Date}} & 
        \multirow{4}{1.1cm}{\centering{Grating}} & 
        \multirow{4}{1.5cm}{\centering{Dispersion (\AA/pix)}} & 
        \multirow{4}{1.5cm}{\centering{Average spectral resolution (km s$^{-1}$)}} & 
        \multirow{4}{1.6cm}{\centering{Wavelength coverage\\ (\AA)}} & 
        \multirow{4}{1.2cm}{\centering{Number of spectra}} & 
        \multirow{4}{1.2cm}{\centering{Exposure time\\(s)}} & 
        \multirow{4}{1.5cm}{\centering{Phase Resolution}} & 
        \multirow{4}{1.4cm}{\centering{Phase coverage}} \\
         
         & & & & & & & \\
         & & & & & & & \\
         & & & & & & & \\
         
        \hline
 		2016-10-05 & G4 & 0.630 & 130 & 3875 -- 5135 
 		           & 22 & 750 & 0.056 & 503.19 -- 504.48  \\
 		2016-10-06 & G4 & 0.630 & 130 & 3875 -- 5135 
 		           & 8 & 750 & 0.056 & 509.85 -- 510.40  \\
 		2016-10-07 & G4 & 0.630 & 130 & 3875 -- 5135 
 		           & 19 & 750 & 0.056 & 516.17 -- 517.53  \\
 		2016-10-09 & G4 & 0.630 & 130 & 3875 -- 5135 
 		           & 15 & 750 & 0.056 & 528.75 -- 529.60  \\ 		           
  		2016-10-10 & G5 & 0.525 & 78 & 5753 -- 6866 
 		           & 27 & 1000 & 0.075 & 535.20 -- 537.27  \\
 		2016-10-11 & G5 & 0.525 & 78 & 5753 -- 6866 
 		           & 10 & 1000 & 0.075 & 546.06 -- 542.75  \\
 		\hline
 	\end{tabular}
 \end{table*}


\section{Analysis}
\label{sec:analysis}

\subsection{Photometry and Orbital Ephemeris}
\label{ssec:a_ephemeris}

\begin{figure}
  \centering
    \includegraphics[trim={0cm 0.5cm 0cm 0cm},width=0.48\textwidth]{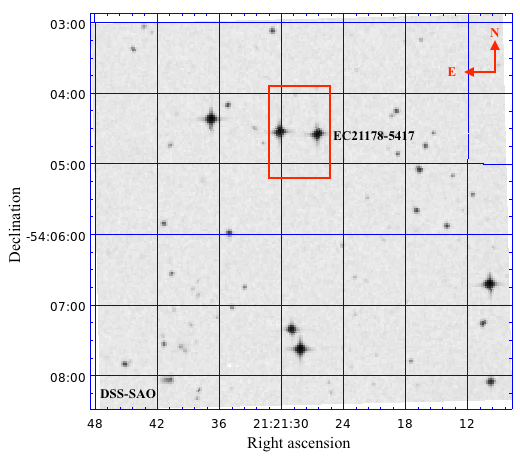}
    \caption{Digital Sky Survey catalogue image of EC2117--54. The field of view of the imager SHOC is represented by the red box, containing the target at the right and the comparison star at the left.}
    \label{fig:fov}
\end{figure}

\begin{figure}
  \centering
    \includegraphics[trim={0cm 0cm 0cm 0cm},width=0.45\textwidth]{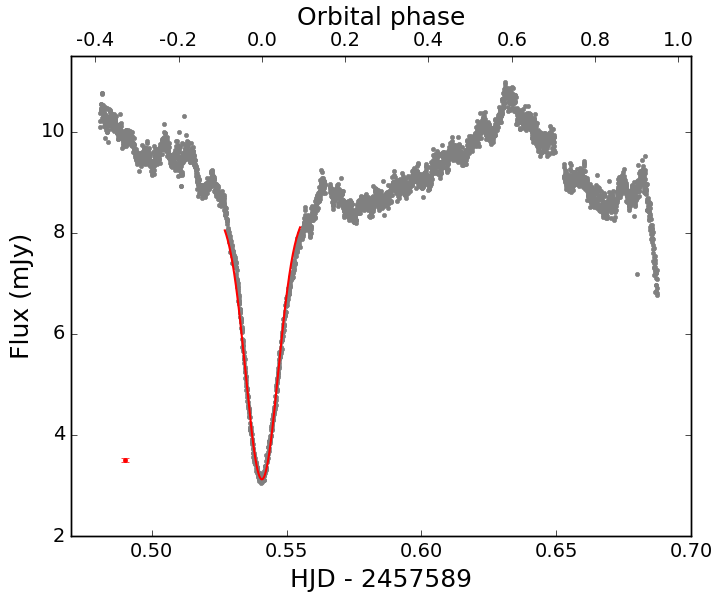}
    \caption{Light curve of EC2117--54 on July 19, 2016. The red dot at the lower left shows the size of the typical error bar on the flux, ignoring systematic errors. The Gaussian fit of the eclipse is represented with the red curve, and the phases derived from the obtained zero phase and orbital period are specified on the upper x-axis.}
    \label{fig:lightcurve}
\end{figure}

Magnitudes were calculated from photometric images relative to the known $r$--band magnitude of the comparison star, i.e. $r = 12.981 \pm 0.003$\, mag (Skymapper Southern Sky Survey, \citealt{2018PASA...35...10W}). A catalogue image of the target and comparison star is presented in Figure~\ref{fig:fov}, with the field-of-view of the SHOC instrument. The comparison star ($\alpha = 21^h 21^m 30.23^s,\; \delta=-54^{\circ} 04' 33.1''$) is also included in the AAVSO Photometric All-Sky Survey \citep{2015AAS...22533616H} and has been observed by {\it Gaia} (\citealt{2016A&A...595A...1G}, \citealt{2018A&A...616A...1G}, \citealt{2018A&A...616A...9L}). Magnitudes were translated into flux units using the flux at $r=0$ \citep{1983ApJ...264..337S}, and the resulting light curve is presented in Figure~\ref{fig:lightcurve}. Magnitudes from SHOC clear filter, calibrated to $r$-band are accurate to 0.1\,mag (e.g. \citealt{2014MNRAS.437..510C}), which translates into systematic errors in flux measurements of approximately $\Delta F \le 0.4\,$mJy around the eclipse. The uncertainties on the differential magnitude and flux with respect to the comparison star are much smaller than the systematic error (see Figure~\ref{fig:lightcurve}).

We fit a Gaussian function to the symmetric, V-shaped signature of the eclipse of the primary star. We use the time of the inferior conjunction to update the zero-point in the ephemeris reported in the literature (\citealt{2003MNRAS.344.1193W}, \citealt{Zietsman:2008tw}, \citealt{Khangale:2013uq}) to assign orbital phases to our spectroscopic observations.

A deeper analysis of the photometric properties of EC2117--54, including a revision of the orbital period, is done in \cite{khangale2019}.

\subsection{Doppler tomography}
\label{ssec:a_doppler}

The spectra of EC2117--54 were prepared for Doppler tomography using the software {\sc molly}. First, orbital phases are assigned to individual spectra (see Section~\ref{ssec:eph}). Spectra at phases around the eclipse, i.e. 0.9 -- 1.1, were excluded as Doppler tomography requires constant visibility of all emission components.  Next, all spectra were normalised by their continuum levels. For every spectral line, the spectra were rebinned in velocity space, and the normalised continuum was subtracted. Finally, the data sets were exported in {\sc ndf} format for Doppler tomography calculations.

\begin{figure*}
  \centering
    \includegraphics[trim={0cm 0cm 0cm 0cm},width=\textwidth]{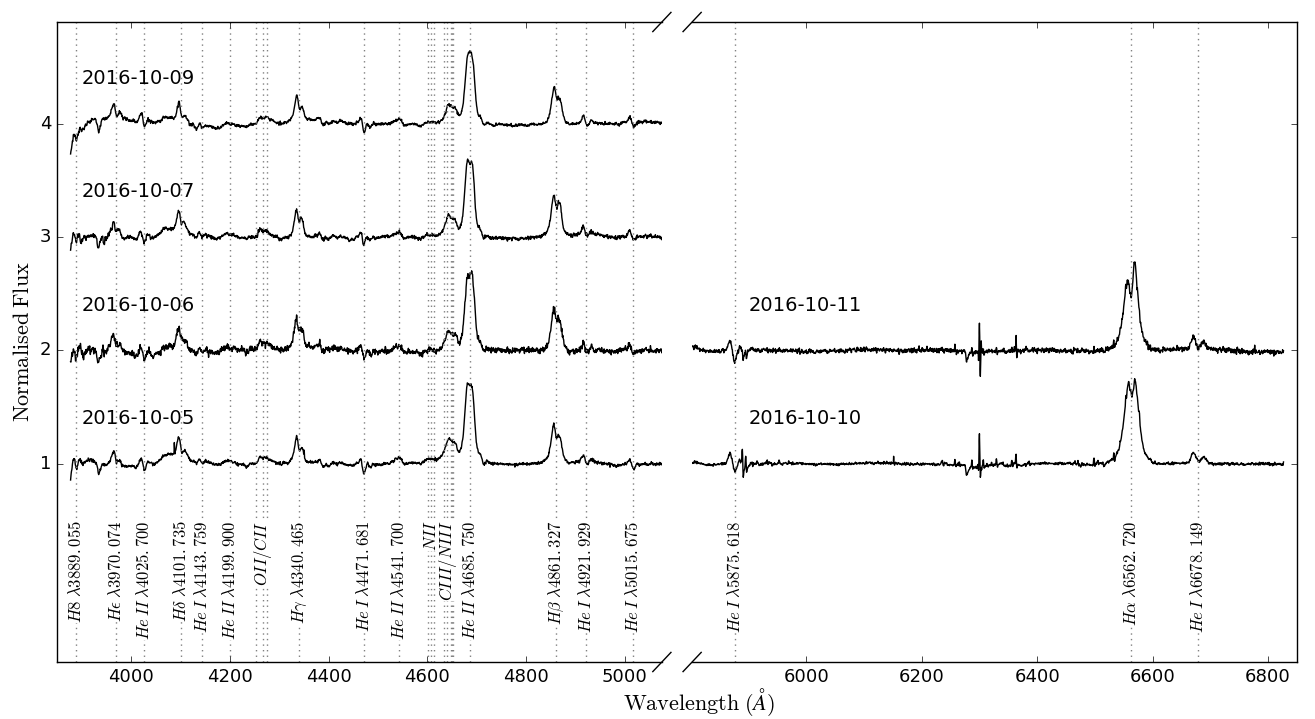}
    \caption{Average spectrum per night normalised by the continuum. Dates and spectral lines are indicated.}
    \label{fig:avspectra}
\end{figure*}

We compute Doppler tomograms for the detected spectral lines with the software {\sc doppler}, that uses maximum-entropy regularization. We followed the procedures outlined in (\citealt{doppler_methods}, hereafter RC19a). The calculations start with a uniform image scaled to the flux of the data. We decreased the $\chi^2$ of the tomograms by 0.5\% in every step of the iterative process. We use a running default image with a Gaussian blurring with a full width at half maximum of 5 resolution elements. Optimal tomograms were selected using the two-dimensional Fourier transform technique outlined in RC19a.

We estimate the systemic velocity, $\gamma$, for every spectral line using three different methods: (i) fitting a double Gaussian profile to the line and subsequently fitting the radial velocity curve, (ii) using the double-Gaussian method and building a diagnostic diagram (\citealt{1980ApJ...238..946S}, \citealt{1985ASSL..113..355S}), and (iii) selecting the value that gives the lowest $\chi^2$ from a series of tomograms at a range of $\gamma$ values, therefore, using Doppler tomography itself. The last method does not assume any specific profile of the lines, and works well for both brighter and fainter lines (RC19a).

We treat the absorption component present in some of the spectral lines by constructing a two-dimensional Gaussian profile in Doppler space centred at $(V_x, V_y)$ = (0,0), with a fixed width of 500 km\,s$^{-1}$ and a peak flux that is just enough to negate the absorption in the data. Then, we translate the Gaussian into data space, matching the spectral and phase resolution of the observed data sets. We add the Gaussian data to the observed data set, and compute the tomogram. We build a different Gaussian in Doppler space to be subtracted from the optimal tomogram, where the peak flux equals the flux at the center of mass of the tomogram (e.g. \citealt{1990ApJ...364..637M}, \citealt{1996A&A...308...97H}).

\subsection{Significance maps}
\label{ssec:a_sigmaps}

We perform a bootstrap routine on the tomograms that allows us to estimate error bars for the flux in the tomogram, and to evaluate the significance of any emitting features in the accretion disc following RC19a., using 2500 bootstrap iterations. For assessing the significance of any feature relative to the accretion disc, the average of flux value $F_{\rm disc}$ and the associated scatter, $\sigma_{\rm disc}$ would be used. Given the low spectral and phase resolution of our data here, the width of the flux distribution across bootstrapped tomograms for any pixel, $\sigma_{\rm{pix}}$, is often larger than the scatter in the flux in the selected accretion disc region. We therefore adopt the conservative approach of selecting the dispersion of the flux in the disc, $\sigma_{F}$, as the maximum value of $\sigma_{\rm{pix}}$ of all pixels in the disc region in the cases that $\sigma_{\rm{disc}} < \sigma_{\rm{pix}}$:
\begin{equation}
    \sigma_{F} = \max\left(\sigma_{\rm{disc}}, \left[\sigma_{\rm{pix}}\right]\mid_{\rm{disc}} \right)
\end{equation}

The significance on the flux is calculated as the highest number, $m$, that verifies the null hypothesis: 
\begin{equation}
  H_0:  F_{\rm{pix}} \ge F_{\rm{disc}} + m \; \sigma_{F}
\end{equation}
It is worth noting that the significance parameter $m$ should not be interpreted as a Gaussian significance, because the distribution of flux in the accretion disc region is not a normal distribution, and we have no knowledge of what the intrinsic characteristics of the parent distribution are. The interpretation of the significance calculated with this method is tailored to the values of $F_{\rm{disc}}$ and $\sigma_{F}$; it is an indicator of the significance of the detected feature standing out with respect to the selected area of the underlying accretion disc.

\subsection{Tomograms in polar coordinates}
\label{ssec:a_polar}

We construct polar tomograms by taking radial crosscuts of the tomogram from the centroid of the emission, i.e. the position of the white dwarf in order to project the tomograms and significance maps onto the radial velocity versus azimuth space, $V_R-\theta$ (\citealt{2001ApJ...551L..89G}, \citealt{detectability_sdw}, hereafter RC19b).

We trace the spiral arms by determining, for every azimuth, at which radial velocity the maximum flux is reached, as projected from the white dwarf: $V_R(F_{\rm{max}})$. We use this parametrisation to derive linear slopes of the spiral arms visible in the tomograms, which are then translated into opening angles using the correlations derived in RC19b.


\section{Results}
\label{sec:results}

\subsection{Light curve and ephemeris}
\label{ssec:eph}

We use the time of the primary eclipse derived from the light curve and the latest orbital period published \citep{khangale2019} to update the zeropoint, leading to the following ephemeris:

\begin{equation}
    \mathrm{HJD} = 2457589.540807 \pm 2 \cdot 10^{-5} + (0.15452724 \pm 10^{-8})\, E
\end{equation}

The difference between the time of this ephemeris and the date of our first spectroscopic observations is 78 days, or approximately 503 orbital periods. The uncertainty on the absolute phase is $\sim$0.003, which translates into an uncertainty of $\sim$1$^{\circ}$ in azimuth. 

\subsection{Spiral structure in EC2117--54}
\label{ssec:tomograms}

\begin{figure*}
  \centering
    \includegraphics[trim={0cm 0cm 0cm 0cm},width=0.93\textwidth]{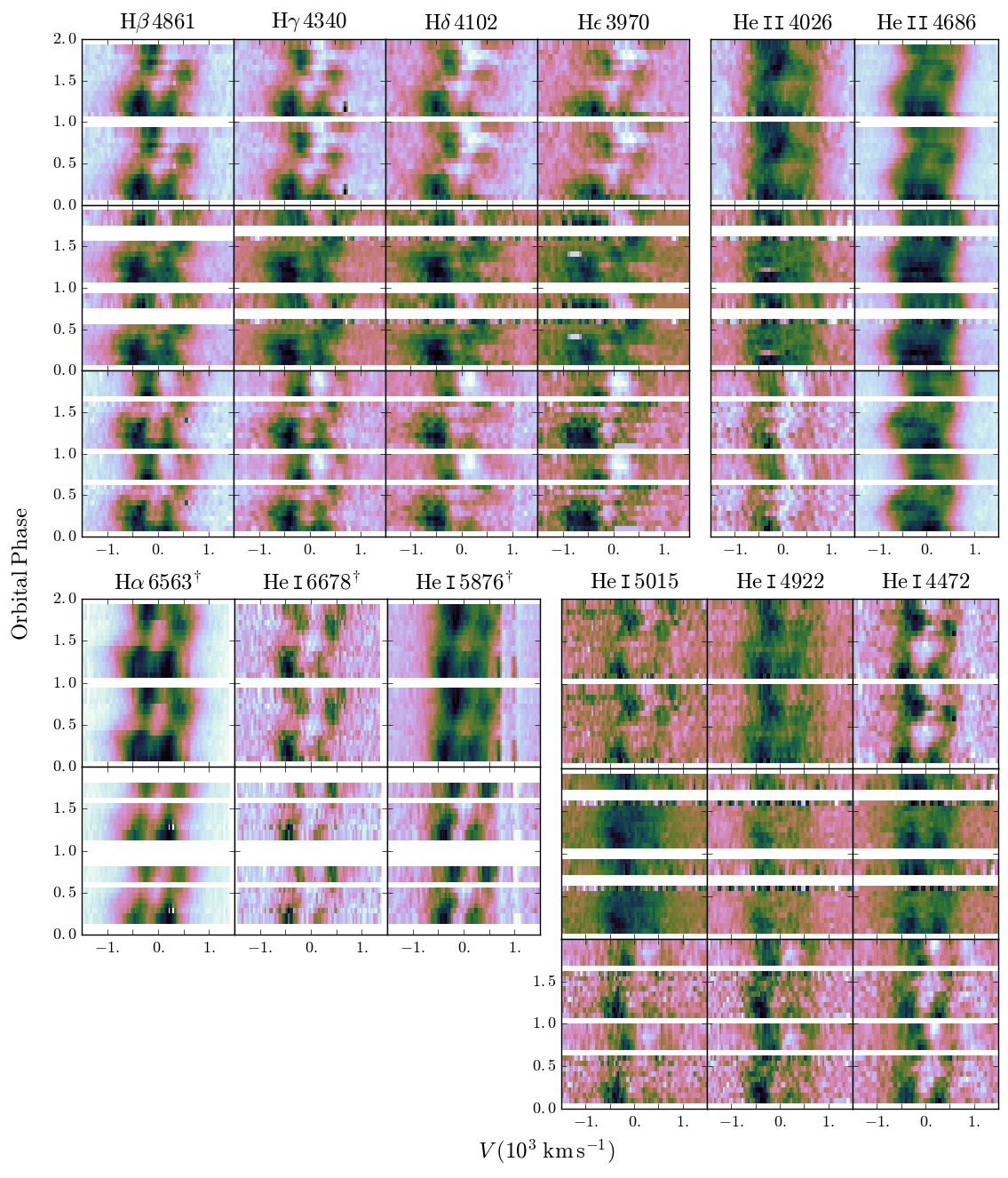}
    \caption{Phase folded trailed spectra for all lines and epochs for which Doppler tomograms are calculated. For lines observed with the blue grating, top panels correspond to epoch 2016-10-05, middle panels to epoch 2016-10-07 and bottom panels to epoch 2016-10-09. $^{\dagger}$For lines observed with the red grating, top panels correspond to epoch 2016-10-10 and bottom panels to epoch 2016-10-11.}
    \label{fig:trailedspectra}
\end{figure*}

\begin{figure*}
  \centering
    \includegraphics[trim={0cm 0cm 0cm 0cm},width=0.93\textwidth]{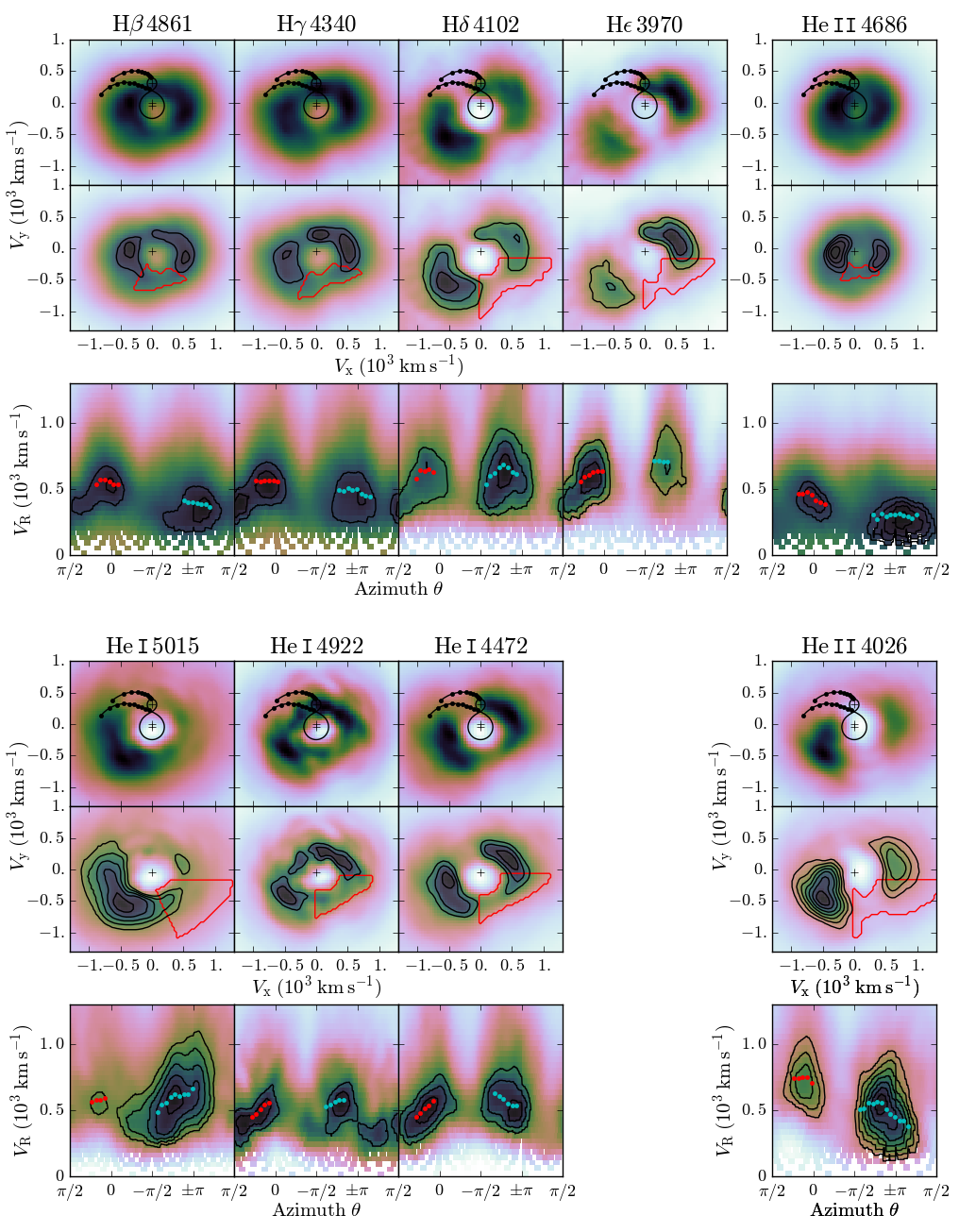}
    \caption{Summary of Doppler tomography on the data on 2016-10-05. For every spectral line, the optimal tomogram, the significance map and the polar tomogram are represented from top to bottom. In the tomograms (top panels), we also represent the centre of mass, the Roche geometry of the system and the ballistic and Keplerian velocities of the gas stream. In the significance maps (middle panels), the black contours indicate the values of the significance parameter, $m$, at increments $\Delta m=1$. The accretion disc region used to define its flux and scatter is represented by the red contour. In the polar tomogram (bottom panels), the significance contours are also represented. The color dots trace the radial velocity of maximum flux of the spiral arms. }
    \label{fig:sum05}
\end{figure*}

\begin{figure*}
  \centering
    \includegraphics[trim={0cm 0cm 0cm 0cm},width=0.98\textwidth]{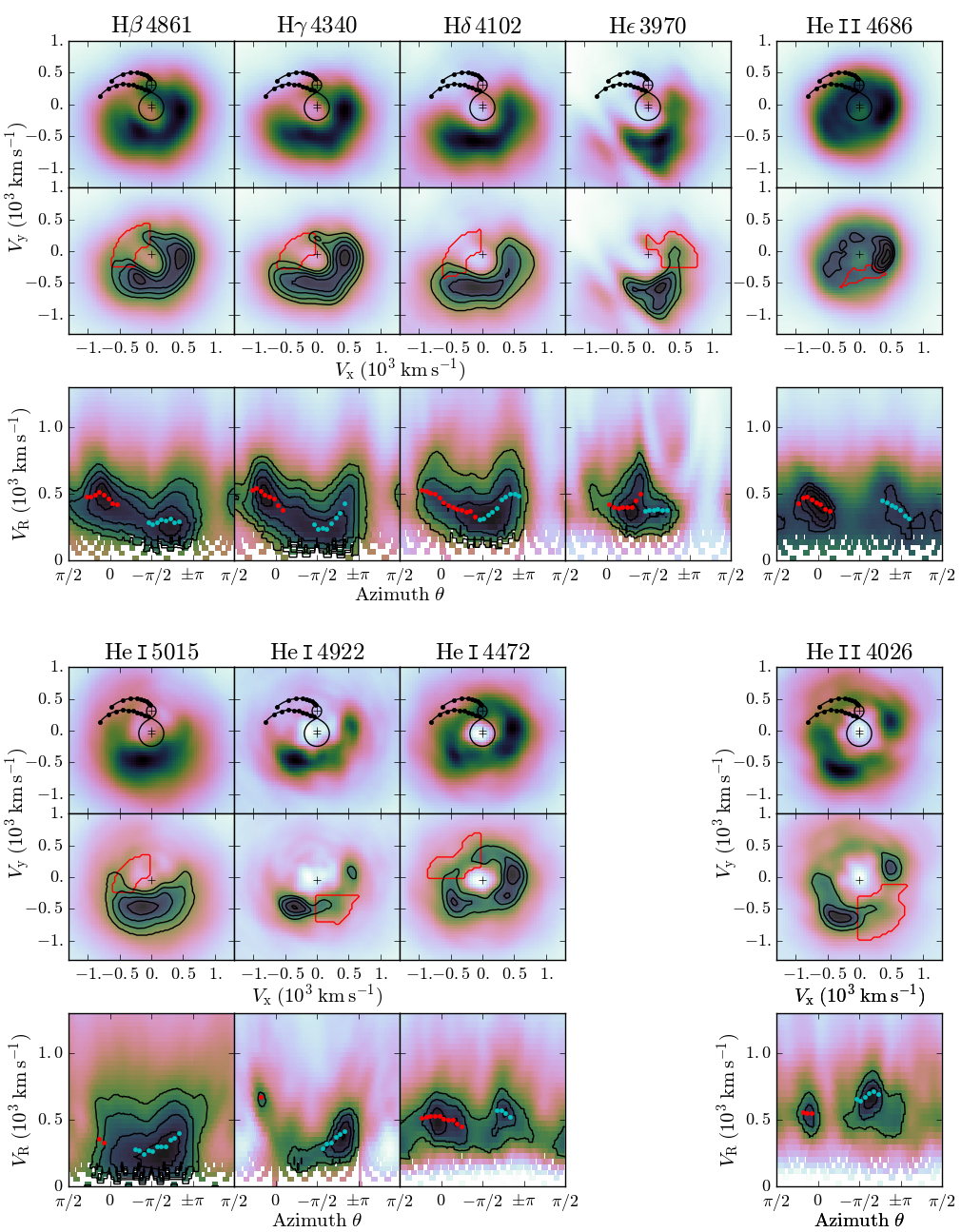}
    \caption{Summary of Doppler tomography on the data on 2016-10-07. For every spectral line, the optimal tomogram, the significance map and the polar tomogram are represented from top to bottom. }
    \label{fig:sum07}
\end{figure*}

\begin{figure*}
  \centering
    \includegraphics[trim={0cm 0cm 0cm 0cm},width=0.98\textwidth]{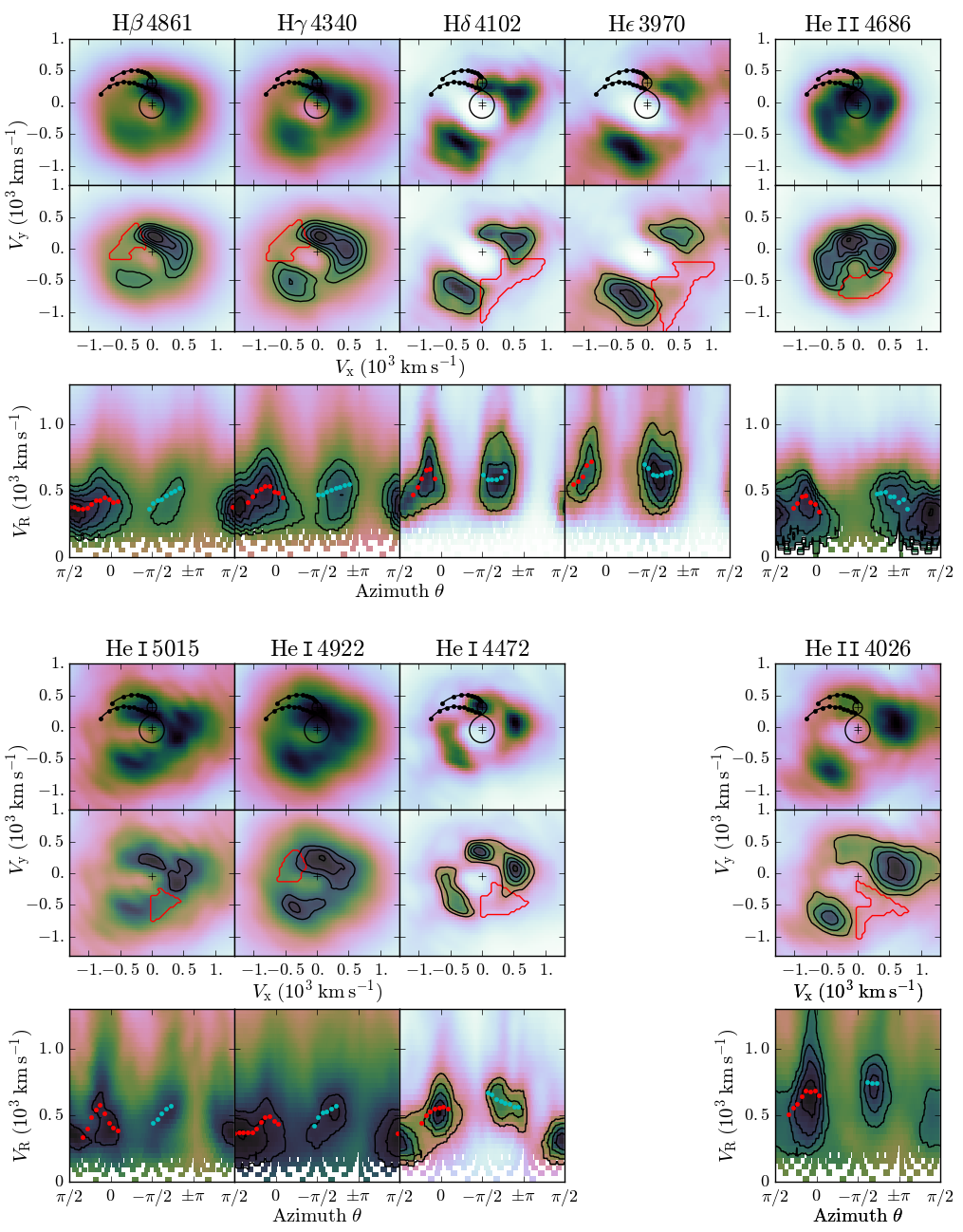}
    \caption{Summary of Doppler tomography on the data on 2016-10-09. For every spectral line, the optimal tomogram, the significance map and the polar tomogram are represented from top to bottom.}
    \label{fig:sum09}
\end{figure*}

\begin{figure*}
  \centering
    \includegraphics[trim={0cm 0cm 0cm 0cm},width=\textwidth]{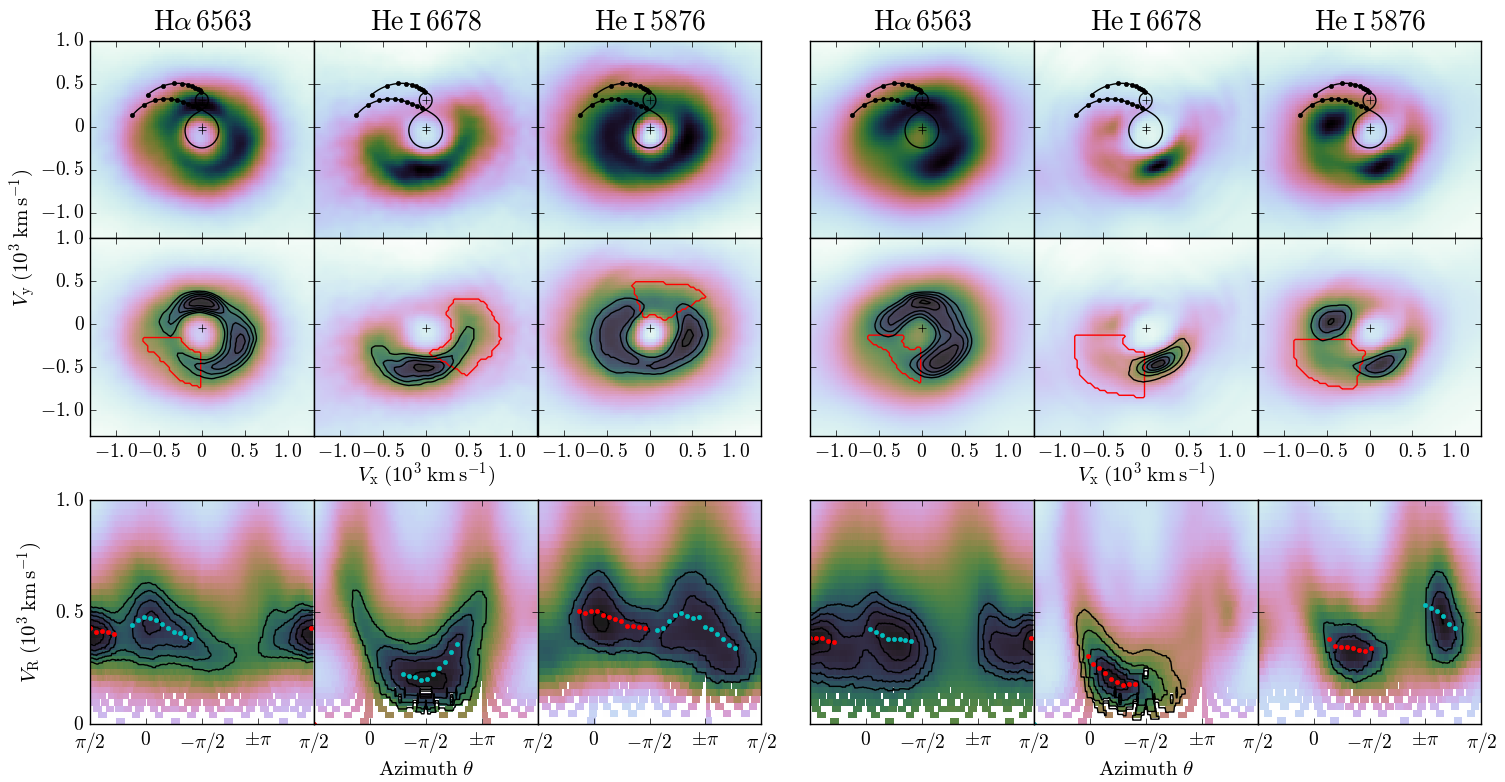}
    \caption{Summary of Doppler tomography on the data on 2016-10-10 and 11. For every spectral line, the optimal tomogram, the significance map and the polar tomogram are represented from top to bottom.}
    \label{fig:sum_red}
\end{figure*}

We present nightly-averaged spectra in Figure~\ref{fig:avspectra}, and the trailed spectra of the spectral lines that we obtained Doppler tomography for in Figure~\ref{fig:trailedspectra}. We represent the results of our study with Doppler tomography in Figures~\ref{fig:sum05}, \ref{fig:sum07} and \ref{fig:sum09} for the three episodes observed with the (blue) grating G4, and in Figure~\ref{fig:sum_red} for the two episodes observed with the (red) grating G5.

The estimates of the systemic velocity by the three methods mentioned in Section~\ref{ssec:a_doppler} agree within the error bars for the brighter lines, i.e. H$\beta$, He\,{\sc ii}\,4686. For fainter lines we rely on the tomographic method. We allow for each line to have its own systemic velocity value rather than selecting a single value for the system to account for possible line-to-line variations reported in other systems, arising for effects other than the system's space velocity, as reported in e.g. the helium system GP Com \citep{1999MNRAS.304..443M}.

The entire set of tomograms shows clear asymmetries in the disc that we identify as spiral structure, in analogy with patterns detected in e.g. IP\,Peg \citep{1997MNRAS.290L..28S} and U\,Gem \citep{2001ApJ...551L..89G}. The significance maps confirm the presence of strong spiral structure standing out against the average accretion disk flux in all tomograms. 

\subsubsection{Blue arm}

Taking the tomograms in the blue arm, on the first night (Figure~\ref{fig:sum05}), as a reference set, the overall distributions of flux, and in particular the spiral structure, present clear similarities for specific lines. For instance, H$\beta$, H$\gamma$, He\,{\sc i}\,4922, He\,{\sc i}\,4472 present two spiral arms of comparable strength: one towards the top right and another towards the bottom left of the tomogram. The He\,{\sc ii} lines exhibit two distinguishable arms as well, but one of them is notably stronger than the other. The exception in the blue arm data on the first night is He\,{\sc i}\,5016 which appears to show only a single-armed spiral feature, with a hint of the second arm. 

Comparing the blue arm data from night to night, we observe clear changes in appearance. The spiral structure in the tomograms of lines such as H$\beta$, H$\gamma$, He\,{\sc i}\,4922, He\,{\sc i}\,4472, He\,{\sc ii}\,4686 and He\,{\sc ii}\,4026 transition from a pattern with two distinct spiral arms on 2016-10-05 to a single-armed structure towards the bottom of the tomogram on 2016-10-07 and then back to a two-armed spiral pattern on 2016-10-09. 

H$\delta$ and H$\epsilon$ show two completely separated arms in the first episode, then a single emission signature on the second epoch, and a two-armed configuration again in the third epoch. In addition, H$\epsilon$ begins with the top right arm as the strongest, as opposite to H$\delta$, but it weakens in favour of the bottom left arm, becoming very similar to H$\delta$. 

For the blue arm data we can conclude that the pattern on the first and third episode, separated by 4 nights, is very similar, but on the second episode, in between the two others, it is very different, indicative of variation on time scales of 1-2 days or shorter.


 \begin{table*}
 	\centering
    \renewcommand*{\arraystretch}{1.3}
 	\caption{Linear slope of the spirals arms as traced by different emission lines, in km s$^{-1}$ deg$^{-1}$.}
 	\label{tab:slope}
 	\begin{tabular}{m{20mm} rrr c rrr} 
 		\hline 
        & \multicolumn{3}{c}{\centering{Top right arm}} &
        & \multicolumn{3}{c}{\centering{Bottom left arm}} \\
        \cline{2-4} 
        \cline{6-8} 
        & 2016-10-05 & 2016-10-07 & 2016-10-09 & 
        & 2016-10-05 & 2016-10-07 & 2016-10-09 \\
        \hline
    
        H$\beta$  &  0.3 $\pm$ 0.4  &  0.9 $\pm$ 0.4  &  --0.6 $\pm$ 0.1  & 
                                     & --0.1 $\pm$ 0.0  & --0.2 $\pm$ 0.2  &  --2.1 $\pm$ 0.1  \\
                                     
        H$\gamma$ &  0.1 $\pm$ 0.1  &  2.3 $\pm$ 0.3  &  --0.4 $\pm$ 0.5  & 
                                     & --0.1 $\pm$ 0.0  & --2.4 $\pm$ 0.6  &  --1.2 $\pm$ 0.1  \\            
        H$\delta$ & --1.1 $\pm$ 0.8  &  1.7 $\pm$ 0.1  &  --3.2 $\pm$ 1.2  & 
                                     & --1.1 $\pm$ 0.7  & --2.6 $\pm$ 0.2  &  --0.6 $\pm$ 0.7  \\ 
                                     
        H$\epsilon$  & --1.5 $\pm$ 0.1  & --1.0 $\pm$ 0.5  & --15.9 $\pm$ 6.5  & 
                                     & --0.1 $\pm$ 0.0  & --0.2 $\pm$ 0.1  &   0.8 $\pm$ 0.5  \\  
                                     
        He {\sc i}\, 5016        & --0.8 $\pm$ 0.2  &  2.8 $\pm$ 3.5  &  --0.2 $\pm$ 1.1  &
                                     &  0.2 $\pm$ 0.2  & --1.3 $\pm$ 0.2  &  --3.4 $\pm$ 0.3  \\ 
                                     
        He {\sc i}\, 4922        & --2.9 $\pm$ 0.2  &  -             &  --1.2 $\pm$ 0.2  &  
                                     & --1.3 $\pm$ 0.2  & --2.6 $\pm$ 0.2  &  --2.7 $\pm$ 0.4  \\ 
                                     
        He {\sc i}\, 4472        & --3.0 $\pm$ 0.1  &  0.7 $\pm$ 0.2  &  --1.7 $\pm$ 0.4  & 
                                     &  2.2 $\pm$ 0.2  &  1.8 $\pm$ 0.5  &   1.7 $\pm$ 0.1  \\ 
                                     
        He {\sc ii}\, 4686       & 0.0 $\pm$ 0.0 & 0.0 $\pm$ 0.0 &  0.0 $\pm$ 0.0 & 
                                     &  0.0 $\pm$ 0.0  & -0.2 $\pm$ 0.1  &  --0.2 $\pm$ 0.1  \\
                                     
        He {\sc ii}\, 4026       &  0.6 $\pm$ 0.5  &  0.2 $\pm$ 0.2  &  --2.3 $\pm$ 0.6  & 
                                     & --0.3 $\pm$ 0.1  & --1.2 $\pm$ 0.4  &   0.1 $\pm$ 0.2  \\
        \hline
        
        & 2016-10-10 & 2016-10-11 &  & 
        & 2016-10-10 & 2016-10-11 &  \\
        \hline
        
        H$\alpha$    &   0.6 $\pm$ 0.1  &  0.5 $\pm$ 0.1    & & 
                                      &   0.9 $\pm$ 0.2  &  0.7 $\pm$ 0.1    & \\
        He {\sc i}\, 6678         &  --1.5 $\pm$ 0.4  &  1.6 $\pm$ 0.2    & & 
                                      &   -             &  -              & \\
        He {\sc i}\, 5876         &   0.8 $\pm$ 0.1  &  0.5 $\pm$ 0.2    & &
                                      &  --0.3 $\pm$ 0.1  &  2.3 $\pm$ 0.1    & \\
        
    \hline
        
 	\end{tabular}
 \end{table*}

\subsubsection{Red arm}

These changes from night to night are also reflected in the red arm data, taken on two consecutive nights, after the blue arm data.

He\,{\sc i}\,5876 shows a two-armed spiral signature in both epochs. The shocks appear somewhat shifted in azimuth with respect to the Balmer lines, i.e., towards the bottom right and the top left of the tomogram. The tomograms of H$\alpha$ are also consistent with this shift, although emission from the secondary star, identifiable as a concentrated spot at the \lq twelve o'clock\rq-position, may be important in this line and appears to merge with the emission from the spiral structure. He\,{\sc i}\,6678 shows a single-armed structure, reflecting that of the He\,{\sc i}\,5015 line in the blue arm.

\subsubsection{Secondary star emission}

The detected emission from the secondary star is especially prominent in the Balmer series, especially for the lower-level transitions. In the first episode it is detectable in H$\beta$, H$\gamma$, and He\,{\sc i}\,4922. The emission from the secondary star is the weakest (relative to the other features) in the second episode, although the significance maps pick up on its signature in He\,{\sc ii}\,4686 and He\,{\sc i}\,4472. The emission is strongest at the third episode, on 2016-10-09, when it is detectable in the majority of the lines. 

In some cases, the emission from the secondary star appears biased towards the trailing side of the star (e.g. He\,{\sc i}\,4922 in Figure~\ref{fig:sum05}, or H$\beta$ and H$\gamma$ in Figure~\ref{fig:sum09}.)

\subsubsection{Opening angles}

We calculate linear slopes of the individual spiral structures as traced by the radial velocity from the white dwarf at maximum flux in the polar tomograms, and present the results in Table~\ref{tab:slope}, following the procedures outlined in RC19b. We observe changes in the absolute values of the estimated linear slopes for different episodes, especially for the Balmer and He\,{\sc i} lines. We also report both positive and negative slopes. A positive slope means that, as we trace the maximum flux of the spiral arm, the radial velocity as seen from the white dwarf increases with azimuth, or decreases with orbital phase. 
In some spectral lines, we see changes from positive to negative slopes as we trace the spiral arm in the polar tomograms, e.g. H$\delta$ on 2016-10-07 and He\,{\sc i}\,6678 on 2016-10-10, and vice versa, e.g. H$\delta$ on 2010-10-05 and He\,{\sc i}\,5876 on 2016-10-10. Changes from positive or negative slope may cancel out since we only derive a single value of the slope for every spiral arm. 

Comparing the derived linear slopes to the correlations between slopes and opening angles given in RC19b, the spiral arms in EC2117--54 can be represented by logarithmic spirals with opening angles, $\phi$, between $-35^{\circ}$ and $-78^{\circ}$, which shows that we cannot constrain the properties of the spiral structure very precisely with current data.

\section{Discussion}
\label{sec:discussion}

The tomographic study of the spectral lines of the system EC2117--54 shows spiral structure in the accretion disc, despite the fact that the data used here are of relatively low spectral and temporal resolution. The first question is therefore if our treatment of the data could artificially introduce these features. The effect of the limited resolution would be to smooth out an underlying spiral pattern rather than introducing these asymmetries artificially. Our treatment of absorption features in the fainter spectral lines by adding an axisymmetric Gaussian profile cannot contribute to the appearance of spiral structure as asymmetries cannot be introduced by symmetric components. 

\subsection{Spiral structure on EC2117-54}

EC2117--54 presents strong asymmetries in its accretion disc, as revealed by the tomograms of several spectral lines. In some cases, for instance the tomograms of He\,{\sc i}\,4922 or He\,{\sc i}\,4472 in Figure~\ref{fig:sum05}, the shape and position of these asymmetries are reminiscent of the structures interpreted as spiral density waves in other systems such as IP\,Peg \citep{1997MNRAS.290L..28S} or U\,Gem \citep{2001ApJ...551L..89G}. In some other cases, for instance the tomograms of He\,{\sc ii}\,4686 in Figure~\ref{fig:sum05}, the spiral structure appears shifted to positions where spiral density waves are not expected based on numerical simulations (e.g. \citealt{1986MNRAS.219...75S}).

Moreover, notable changes in the spiral structure are observed from night to night, or every two nights in the case of the bluer lines in a consistent way for the spectral lines studied. These shifts are at odds with the spiral arms detected in the dwarf nova U\,Gem \citep{2001ApJ...551L..89G}, that remain in the same position throughout three epochs during the decline of an outburst, although only He\,{\sc ii}\,4686 is explored in U\,Gem. Direct comparison of the systems EC2117--54 and U\,Gem may not fully applicable, since the pattern in U\,Gem is there only during outburst, and also as a consequence of the dwarf nova outburst. In EC2117--54 the disc is in a quasi-steady state, not undergoing a disc instability event as in U\,Gem. In a theoretical context, stationarity is assumed (e.g. \citealt{1987A&A...184..173S}) based on the quasi-stationary spiral waves found by \citealt{1986MNRAS.219...75S}, which reduces the complexity of the analytical problem. 

The location and fast evolution of the spiral structure pose problems on the interpretation of the asymmetries in EC2117--54 as spiral density waves. This favours alternative scenarios and interpretations of Doppler tomograms. \citet{2001AcA....51..295S} suggested that the asymmetric emission is related to enhanced surface density areas due to tidally perturbed plasma orbits. The observed uneven relative flux of the spiral arms can be explained in this framework. However, the intensified emission appears at the same position where spiral waves are expected, and therefore cannot account for our observations. \citet{2002MNRAS.330..937O} calculated a three-dimensional, semi-analytical model of a tidally distorted accretion disc and proposed that the emission observed in the tomograms comes from irradiated, vertically thickened sector in the outer regions of the disc. In this framework, positional shifts of the asymmetries take place with changes to the effective viscosity $\alpha$ or the Weissenberg number (defined as the relaxation time multiplied by the angular velocity), which can explain our observations. As noted by \citet{2002MNRAS.330..937O}, however, a realistic treatment of the gas inflow into the disc and transport of mass through the disc needs to be included in the theoretical treatment. Numerical simulations of the incoming gas stream show that vertical bulges develop at several phases, i.e. $\phi$ = 0.2, 0.5 and 0.8, reaching scale heights of 10--20\% on average at the outer parts of the disc (\citealt{1991PASJ...43..809H}, \citealt{1993MNRAS.264..691M}). These vertically thickened regions of the disc can lead to enhanced emission when irradiated and also to relevant self-obscuration of certain regions of the disc, given the orbital inclination of EC2117-54, which can explain the structures we observe in the tomograms. The vertical extent of such bulges depends strongly on the mass accretion rate, disappearing when mass inflow is artificially turned off \citep{1991PASJ...43..809H}. Considering the timescales in which hydrostatic equilibrium is stablished, vertical structures like this would live for time scales comparable to the rotation period \citep{1995CAS....28.....W}, which is of the order of hours.

The radial extension of the disc in the tomogram is dominated by the spiral structure, and appears similar in the tomograms of all spectral lines considering the spectral resolution of the data. If the observed emission is produced mainly in structures limited in space, i.e. spiral waves or vertically enhanced regions, then a similar radial extent is expected. If the observed asymmetries result from self-obscuration this does not need to be true, and the optical depth for different phases and lines needs to be taken into account.

The linear slopes of the spiral arms are indicative of how openly wound the spiral waves are (RC19b), which in turn is indicative of the temperature of the disc \citep{1999MNRAS.307...99S}. However, our results do not show a clear trend across different episodes, and the changes are not consistent for different lines. Our calculated slopes are limited to a phase resolution of about 20$^{\circ}$ while the spiral arms in the tomograms are traceable for less than 90$^{\circ}$. In some cases, we may be tracking the underlying accretion disc. Finally, a single value of the inclination of the spiral shocks is not appropriate in cases where slopes reverse arithmetic sign along the same spiral arm.

EC2117--54 (\citealt{khangale2019}, this work), U\,Gem \citep{2001ApJ...551L..89G} and IP\,Peg (e.g. \citealt{2000MNRAS.313..454M}) are the only systems for which observational studies on the evolution of spiral patterns at short time scales exist. Additional observational and theoretical efforts are necessary to understand the origin of asymmetric discs such as the observed here.

\subsection{Comparison to other novalikes}

Novalikes that exhibit spiral structure in their Doppler maps are rare. V347\,Pup shows a two-armed spiral structure with emission from the secondary star in the Balmer lines \citep{1998MNRAS.299..545S}. The overall appearance of the spiral pattern in V347\,Pup is very similar to our Balmer tomograms on the episode 2016-10-09. The long-term persistence of the spiral structure in V347\,Pup seems uncertain, as \citet{2001LNP...573...45S} and \citet{2005MNRAS.357..881T} reported two separate observational studies years later in which the asymmetries in the disc had vanished in the Balmer lines, but were visible in He\,{\sc i} lines instead. In this sense, the spiral structure in EC2117--54 might be more stable over secular changes as Doppler maps on 2010-10-09 closely resemble the preliminary tomograms presented in \citet{Khangale:2013uq}, perhaps with slight differences in H$\alpha$. 

V3885\,Sgr also shows spiral waves in H$\beta$, H$\gamma$, and with less strength, in He\,{\sc i}\,4472 \citep{2005MNRAS.363..285H}, very similar to those in U\,Gem and comparable to our tomograms in the 2016-10-05 episode. 

The spiral pattern in UX\,UMa \citep{2011MNRAS.410..963N}, only detected in H$\alpha$, presents a two-armed structure but the arm typically at the top right of the tomogram is shifted towards later phases, similarly to our He\,{\sc i}\,4472 tomogram in the 2016-10-07 episode. This suggests that the differences observed from system to system are related to distinct states of their accretion discs during data acquisition. 

\subsection{Emission from the secondary star}

The emission from the secondary star is detectable on 2016-10-05, and in particular on 2016-10-09. The tomograms in these two episodes are remarkably similar, while they differ from the results on 2016-10-07. The detection of emission from the secondary star may point to higher levels of irradiation due to higher temperatures in the disc, or, when irradiation is absent, the secondary star is perhaps shielded by a flared-up accretion disc or vertically enhanced structures at the outer rim of the disc (\citealt{1991PASJ...43..809H}, \citealt{2002MNRAS.330..937O}).

In many cases, the emission from the secondary star does not appear at the \lq twelve-o' clock\rq\, position, but is shifted towards the trailing side of the star. The uncertainties in the orbital ephemeris derived in Section~\ref{ssec:eph} are too small to explain this shift. This is also observed in the tomograms of V347\,Pup (\citealt{1998MNRAS.299..545S}, \citet{2001LNP...573...45S}, \citet{2005MNRAS.357..881T}) and UX\,UMa \citep{2011MNRAS.410..963N}. \citet{1998MNRAS.299..545S} speculate that if the gas stream spills over the impact region with the disc, it could provide an optically thick material that would shield the leading side of the secondary star from irradiation. This, however, requires that the bright spot or the stream have a structure that extends over several scale heights. Moreover, there are no indications of the bright spot in the Doppler maps of these studies, although this is somewhat expected because evidence of a bright spot can hinder the detection of asymmetries in the disc \citep{2000A&A...356L..33J}. Our observational results do not show any evidence of a bright spot either, perhaps favouring a scenario where the spiral structure observed in EC2117--54 is related to self-obscuration by vertical enhancements in the disc that can protect the secondary star from irradiation (\citealt{1991PASJ...43..809H}, \citealt{2002MNRAS.330..937O}).

\section{Conclusions}
\label{sec:conclusions}

We have reported on the detection of time-variable spiral structure in the accretion disc of the novalike EC2117--54. The spiral structure is observed in the tomograms of all the spectral lines studied, but it is especially strong in H$\beta$, H$\gamma$, He\,{\sc i}\,4472 and He\,{\sc ii}\,4686. The spiral pattern is persistent over a week, although there are night-to-night changes in the position and strength of the spiral arms. 

EC2117--54 joins a small group of peculiar novalikes that exhibit spiral structure in their Doppler tomograms, after V347\,Pup, V3385\,Sgr and UX\,UMa. The Doppler maps published for these other novalikes resemble our tomograms at particular episodes, but an analogous observational study of the night-to-night evolution of the emission pattern is yet to be done. 

The location and evolution of the spiral structure challenges the interpretation of these asymmetries as the signature of spiral density waves, as the latter are expected to be stationary at specific positions. Other scenarios, such as tidally thickened sectors of the disc or vertically enhanced bulges from the interaction of the gas stream with the disc, can explain our observations.


\section*{Acknowledgements}

This work is part of the research programme with project number 614.001.207, which is financed by the Netherlands Organisation for Scientific Research (NWO). 

This research was supported in part by the National Science Foundation (NSF) under Grant No. NSF PHY17-48958. PAW and ZNK acknowledges the financial support from the NRF and the University of Cape Town. PJG acknowledges support from the NRF under grant 111692. RRC and PJG also thank the University of Cape Town for their hospitality, during two visits supported by the NWO-NRF Bilateral Agreement in Astronomy. RRC and PJG thank the Kavli Institute for Theoretical Physics at the University of California, Santa Barbara for valuable scientific exchange during the programme DISKS17 from January to March 2017. 

RRC thanks the South African Astronomical Observatory, Lisa Crause and Kerry Paterson for a very successful observing run. RRC also thanks D. Steeghs for useful comments on Doppler tomography.

The Digitized Sky Surveys were produced at the Space Telescope Science Institute under U.S.  Government grant NAG W-2166. The images of these surveys are based on photographic data obtained using the Oschin Schmidt Telescope on Palomar Mountain and the UK Schmidt Telescope. The plates were processed into the present compressed digital form with the permission of these institutions. The Oschin Schmidt Telescope is operated by the California Institute of Technology and Palomar Observatory. The UK Schmidt Telescope was operated by the Royal Observatory Edinburgh, with funding from the UK Science and Engineering Research Council (later the UK Particle Physics and Astronomy Research Council), until 1988 June, and thereafter by the Anglo-Australian Observatory. The blue plates of the southern Sky Atlas and its Equatorial Extension (together known as the SERC-J), as well as the Equatorial Red (ER), and the Second Epoch [red] Survey (SES) were all taken with the UK Schmidt.

This work has made use of data from the European Space Agency (ESA) space mission Gaia. Gaia data are being processed by the Gaia Data Processing and Analysis Consortium (DPAC). Funding for the DPAC is provided by national institutions, in particular the institutions participating in the Gaia MultiLateral Agreement (MLA). The Gaia mission website is \url{https://www.cosmos.esa.int/gaia}. The Gaia archive website is \url{https://archives.esac.esa.int/gaia}.

This research has made use of the VizieR catalogue access tool, CDS, Strasbourg, France (DOI: 10.26093/cds/vizier). The original description of the VizieR service was published in A\&AS 143, 23

{\sc iraf} is distributed by the National Optical Astronomy Observatories, which are operated by the Association of Universities for Research in Astronomy, Inc., under cooperative agreement with the National Science Foundation \citep{1993ASPC...52..173T}. {\sc p}y{\sc raf} and {\sc p}y{\sc fits} are products of the Space Telescope Science Institute, which is operated by AURA for NASA. 

We have used the software {\sc molly} and {\sc doppler} by T. Marsh for preparing spectra and computing Doppler tomograms, publicly available on the website \url{http://deneb.astro.warwick.ac.uk/phsaap/software/}.

We have made extensive use of {\sc python} packages: {\sc numpy} \citep{Oliphant:2006wm}, {\sc matplotlib} \citep{Hunter:2007}, {\sc scipy} \citep{2001..scipy}, and {\sc astropy} \citep{2013A&A...558A..33A}. We have also use the package {\sc starlink-pyndf} by T. Jenesses, G. Bell and T. Marsh, which is publicly available on the website \url{https://github.com/timj/starlink-pyndf}.




\bibliographystyle{mnras}
\bibliography{ec2117-54}  







\bsp	
\label{lastpage}
\end{document}